\documentclass[iop]{emulateapj}

\begin{document}

\title{Similarity of the Optical-IR and $\gamma$-ray Time Variability of Fermi Blazars}

\author{Ritaban Chatterjee\altaffilmark{1}, C. D. Bailyn\altaffilmark{1}, E. W. Bonning\altaffilmark{2}, M. Buxton\altaffilmark{1}, P. Coppi\altaffilmark{1}, G. Fossati\altaffilmark{3}, J. Isler\altaffilmark{1}, L. Maraschi\altaffilmark{4}, C. M. Urry\altaffilmark{2}}

\altaffiltext{1}{Department of Astronomy, Yale University, PO Box 208101, New Haven, CT 06520-8101; ritaban.chatterjee@yale.edu.}
\altaffiltext{2}{Department of Physics and Yale Center for Astronomy and Astrophysics, Yale University, PO Box 208121, New Haven, CT 06520-8121.}
\altaffiltext{3}{Department of Physics and Astronomy, Rice University, 6100 Main St., Houston, TX 77005.}
\altaffiltext{4}{INAF--Osservatorio Astronomico di Brera, V. Brera 28, I-20100 Milano, Italy.}

\begin{abstract}
We present the time variability properties of a sample of six blazars, AO 0235+164, 3C 273, 3C 279, PKS 1510-089, PKS 2155-304, and 3C 454.3, at optical-IR frequencies as well as $\gamma$-ray energies. These observations were carried out as a part of the Yale/SMARTS program during 2008-2010 that has followed the variations in emission of the bright Fermi-LAT-monitored blazars in the southern sky with closely-spaced observations at BVRJK bands. We find that the optical-near IR variability properties are remarkably similar to those at the $\gamma$-ray energies. The discrete auto-correlation functions of the variability of these six blazars at optical-IR and $\gamma$-ray energies do not show any periodicity or characteristic timescale. The power spectral density (PSD) functions of the R-band variability of all six blazars are fit well by simple power-law functions with negative slope such that there is higher amplitude variability on longer timescales. No clear break is identified in the PSD of any of the sources. The average slope of the PSD of R-band variability of these blazars is similar to what was found by the Fermi team for the $\gamma$-ray variability of a larger sample of bright blazars. This is consistent with leptonic models where the optical-IR and $\gamma$-ray emission is generated by the same population of electrons through synchrotron and inverse-Compton processes, respectively. The prominent flares present in the optical-IR as well as the $\gamma$-ray light curves of these blazars are predominantly symmetric, i.e., have similar rise and decay timescales, indicating that the long-term variability is dominated by the crossing time of radiation or a disturbance through the emission region rather than by the acceleration or energy-loss timescales of the radiating electrons. For the blazar 3C 454.3, which has the highest-quality light curves, the total energy output, the ratio of $\gamma$-ray to optical energy output, and the $\gamma$-ray vs. optical flux relation differ in the six individual flares observed between 2009 August and December. The results are consistent with the location of a large $\gamma$-ray outburst in 3C 454.3 during 2009 December being in the jet at $\sim$18 pc from the central engine. This poses strong constraints on the models of high energy emission in the jets of blazars.
\end{abstract}


\section{Introduction}
Blazars are a sub-class of active galactic nuclei (AGN) characterized by a prominent jet pointing within a few degrees of our line of sight \citep{urr95}. In many cases, the jets are luminous over a wide range of wavelengths from radio to $\gamma$-rays. Due to the proximity of the jet axis to the line of sight, emission from the jet is relativistically beamed and hence amplified by an order of magnitude or more for many blazars. The observed timescales are also shorter than that in the rest frame of the jet plasma. Most of the observed radio to optical (and in some cases X-ray) emission from the blazars is due to synchrotron radiation in the jet \citep{bre81,urr82,imp88,mar98}. X-rays and $\gamma$-rays may be due to inverse Compton scattering by the same energetic electrons radiating synchrotron emission \citep[the so called leptonic models; e.g.,][]{bot07} or due to synchrotron radiation by protons co-accelerated with the electrons in the jet, interactions of these highly relativistic protons with external radiation fields, or proton-induced particle cascades \citep[hadronic models; e.g.][]{muc01,muc03}. 

In the leptonic model, the source of the seed photons that are being scattered may be the synchrotron photons generated within the jet, in which case it is termed synchrotron self-Compton (SSC) process \citep{mar92,chi02,arb05}, or from outside the jet (radiation from broad emission line region or BLR, accretion disk, or dusty torus), termed external Compton or EC process \citep{sik94,cop99,bla00,der09}. 

Time variability across multiple wavebands is a defining property of blazars and has been used to probe the location and physical processes related to the emission at very fine resolutions \citep[e.g.,][]{cha08,mar08,mar10,jor10,agu11}. Most of the observed radiant energy produced in many blazars peaks in the $\gamma$-ray part of the spectrum. But until recently, long-term and well sampled $\gamma$-ray light curves of blazars were not available. Expansion of such monitoring to a wide range of $\gamma$-ray energies (20 MeV to 300 GeV) is now occurring through the \textit{Fermi Gamma-Ray Space Telescope} which was launched in 2008 \citep{abd10_timing}. The Fermi bright source list from the first 11 months contains more than 600 \citep{abd10_catalog} and the \textit{Fermi} 2-yr catalog \citep{abd11_2yrcatalog} contains more than 1000 AGNs most of which are blazars. 

Investigating the time variability of the data from Fermi and supporting multi-frequency monitoring with statistically robust techniques is one of the most promising ways to understanding how and where the multi-wavelength emission is produced in the jets of blazars. For example, in the leptonic models, the low and high-energy variability should be correlated while in some versions of the hadronic models, this correlation may be weak or absent. For this reason, multi-frequency monitoring programs are crucially important. 

Here we report on the Yale/SMARTS optical--near IR monitoring program\footnote{http://www.astro.yale.edu/smarts/fermi/} during 2008-2010 which has followed the variations in emission of the Fermi-LAT monitored blazars in the southern sky with closely-spaced observations \citep[see][for details of data acquisition and calibration]{bon11}. Six blazars were detected for a sufficiently long interval by the Fermi-LAT during 2008-10: AO 0235+164, 3C 273, 3C 279, PKS 1510-089, PKS 2155-304, and 3C 454.3. In order to quantify their variability properties, we calculate the power spectral density and the auto-correlation function, and carry out flare decomposition analysis of the SMARTS light curves of these six sources. We then compare the results with similar properties of the $\gamma$-ray variability as discussed by \citet{abd10_timing} who analyzed 11-month-long GeV light curves of a sample of bright blazars from the Fermi 3-month source catalog. Possible models of emission may be distinguished by comparing the characteristics of the flux variations at lower (e.g. optical-IR) and higher (e.g. $\gamma$-ray) energies. 

The auto-correlation function (ACF) can provide insights into the nature of variability. \citet{abd10_timing} have shown that the ACFs of the $\gamma$-ray variability of a sample of bright blazars do not show any hint of quasi-periodicity or characteristic timescale. Here we calculate the ACFs of those six blazars to search for such a timescale in optical or near IR bands. Comparison of the ACF of the optical-IR variability with that of the $\gamma$-ray variability can be used to understand the relation between the mechanism and location of emission in these wave bands. For example, in the leptonic models, the optical-IR and $\gamma$-ray emission is generated by the same population of electrons. Hence, it is expected that the width and shape of the ACF will be very similar at both energies. However, some hadronic models are also able to reproduce this behavior and hence we cannot exclude those models on this basis. In this paper, we calculate the ACFs of the $\gamma$-ray variability of those blazars as well to facilitate the comparison of the $\gamma$-ray ACFs to those at optical-IR wave bands. 

Power spectral density (PSD) analysis is a common technique to characterize time variability. The PSD corresponds to the power in the variability of emission as a function of timescale. \citet{abd10_timing} show that the PSDs of the $\gamma$-ray light curves of a sample of bright blazars are well-described by simple power laws, corresponding to ``red noise". Red noise is defined as uncorrelated fluctuations where power density decreases with increasing frequency. In the case of an astronomical time series this translates to having larger amplitude variations at longer timescales. X-ray PSDs of Seyfert galaxies and black hole X-ray binaries (BHXRBs) can be fit by a piece-wise power law, with a slope that steepens above a ``break-frequency". The break-frequency is related to a characteristic timescale (the origin of which is not well-understood) that is proportional to the mass of the central black hole \citep{now99,utt02,mch04,mar03,pou01,ede99,cha09,cha11}. We calculate the PSD of the optical variation in these blazars in order to search for the existence of such a characteristic timescale. Similar to the ACFs, if the optical and $\gamma$-ray emission is radiated by the same electrons, the shape and slope of the PSD functions at these two energy-bands are expected to be similar. 

Where well-sampled simultaneous light curves exist, we can compare the properties of the individual $\gamma$-ray and optical-IR outbursts to investigate the nature of the emission mechanisms. The total energy output of an optical-IR flare can also be compared to that of the corresponding $\gamma$-ray flare. To achieve this, we decompose the light curves into individual (sometimes overlapping) flares, similar to \citet{val99}, each with exponential rise and decay. \citet{abd10_timing} have also analyzed segments of Fermi light curves of 10 sources to investigate the properties of individual flares. They found that most of the outbursts are symmetric in nature. In this paper, we use a slightly different algorithm to perform a similar analysis of the SMARTS R-band and Fermi light curves of the six blazars in our sample.

There are two main classes of blazars, namely, flat spectrum radio quasars (FSRQs) and BL Lacertae type objects (BL Lacs). One of the main differences in the observable properties of these two classes is that the BL Lacs have much weaker or no observable emission lines. This is probably due to different parent populations of this two classes, i.e., FSRQs are linked with FR II and BL Lacs are with FR I radio galaxies \citep{urr95}. Although the bright jet continuum emission complicates our view of the unbeamed thermal components \citep{geo98}, and it has caused some debate about the meaning of the BL\, Lac--FSRQ difference \citep{ver95,cor00,rai07}, it is now apparent that the majority of BL\,Lac objects have intrinsically less luminous thermal emission, pointing at some fundamental difference \citep{urr95,fos10,ghi11_transition}. A change in the properties of their central regions associated with a critical value of the mass accretion rate (in Eddington units) has been proposed  \citep{ghi09_divide}, and there are indications that it might hold for radio-galaxies, too \citep{ghi10,chiaberge02}.

In addition to the FSRQ/BL Lac divide, the frequency of the synchrotron peak (in $\nu F_\nu$ vs. $\nu$ diagrams) has emerged as one of the most important observational distinctions, leading to the classification of blazars based on its value, namely, Low/Intermediate/High Synchrotron Peak objects (LSP/ISP/HSP). \citet{fos98} showed that blazar SEDs seem to change systematically with luminosity: the most powerful objects are LSP, while HSP SEDs are associated with relatively weak sources. This result has been supported by studies of high-z blazars, low power objects and recently of $\gamma$-ray selected samples \citep{fab01_1,fab01,cos01,ghi09,sam10}, but more recent work suggests that the hypothesized relationship between SED peak and luminosity may be more complex than a simple sequence \citep{mey11}.

Among the six blazars in our sample, PKS 2155-304 is a high-synchrotron-peaked BL Lac type object and the rest are FSRQs as well as low-synchrotron-peaked objects. Here we test if any significant difference exists between the multi-wave band variability properties of the FSRQs in our sample and that of PKS 2155-304.

In {\S}2 we calculate the auto-correlation functions of the optical, infrared, and $\gamma$-ray light curves of our sample of six blazars while in {\S}3 we present the power spectral analysis, its results and implications. In {\S}4 we determine the shortest variability timescales present in our data. We model the flares in optical and $\gamma$-ray light curves and discuss the properties of the flares in {\S}5 while in {\S}6 we compare the time variability properties of FSRQs and BL Lacs. In {\S}7 we decompose a segment each of contemporaneous Fermi and SMARTS light curves of the blazar 3C 454.3 to investigate the nature of the outbursts and the corresponding emission processes. {\S}8 presents the summary and conclusions. The $\gamma$-ray light curves analyzed in this work are at 0.1-300 GeV energy band and are taken from the \textit{Fermi} LAT table of monitored sources provided by the \textit{Fermi} team.

\section{Auto-Correlation Function}
\begin{center}
\begin{deluxetable}{cccc}
\tablecolumns{4} 
\tablewidth{0pt}
\tablecaption{The half width at half maximum (HWHM) of the auto-correlation functions of the blazars at three wave bands.\label{acf_table}}
\tablehead{
\colhead{} &\multicolumn{3}{c}{HWHM (Days)} \\
\cline{2-4} \\ 
\colhead{Object}& \colhead{B-Band} & \colhead{J-Band}	& \colhead{$\gamma$-ray}}
\startdata
AO 0235+164	&	24	&	24	&	26	\\
3C 273		&	14	&	13	&	8	\\	
3C 279		&	41	&	28	&	8	\\	
PKS 1510-089	&	8	&	12	&	5	\\	
PKS 2155-304	&	27	&	31	&	11	\\
3C 454.3		&	9	&	9	&	9
\enddata
\end{deluxetable}
\end{center}
The discrete cross-correlation function (DCCF) is a method to calculate the cross-correlation of unevenly sampled discrete data \citep{ede88}. We use this to calculate the discrete auto-correlation function (DACF) of the variability of the blazars in our sample at $\gamma$-ray energies, as well as B and J-band frequencies. The width of the ACF may be related to a characteristic size-scale of the corresponding emission region, while equally spaced and repeated peaks or drops in the function shape can point out characteristic timescales and provide hints of possible quasi-periodicity. 

The optical-IR light curves obtained by SMARTS have seasonal gaps ($\sim$3-4 months) due to the source going behind the Sun. These large gaps distort the ACF. To avoid this, for each of the six blazars, we select the longest segment without such large gaps and use them to calculate the ACFs. The length of such segments were 200$-$250 days for all six blazars in our sample. This enables us to avoid distortion due to large data-gaps while keeping the resultant ACF a significant representative of the nature of the entire light curve. The average sampling rate of the SMARTS light curves was one data point every 2-3 days. The light curves from Fermi do not contain the Sun-gaps but the blazar may not be consistently detected during the selected segments of the optical-IR light curves. Therefore, while making the selection of the optical-IR segments, we chose those intervals when the blazar was detected by Fermi at least once in 2-3 days on average so that the $\gamma$-ray sampling is similar or better than that in the optical-IR bands. We chose the B and the J-band light curves for comparison with the $\gamma$-rays since those are the bands with the widest separation among our data while the K-band data have a sampling frequency less than once in 2-3 days in some cases.

\begin{figure*}
\includegraphics[height=7in,width=2in,angle=-90]{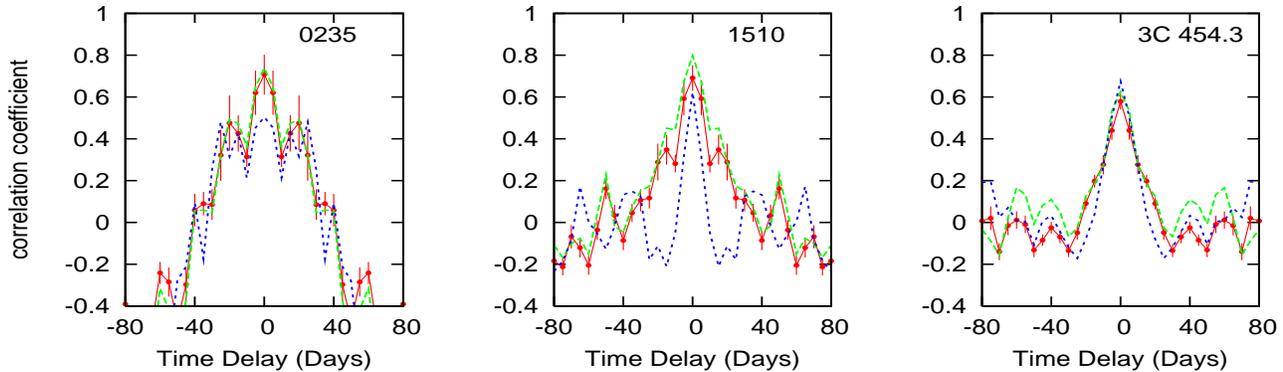}
\caption{Discrete auto-correlation functions of the $\gamma$-ray (blue dotted line), $B$-band (red solid line and filled circles with error bars), and $J$-band (green dashed line) light curves of AO 0235+164, PKS 1510-089, and 3C 454.3. Uncertainties shown by the red data points are characteristic of other bands as well. In these three sources, the temporal sampling in all 3 bands is the same, hence this is a comparison of the intrinsic variability properties of these blazars.}
\label{acf1}
\end{figure*}

The ACF of a light curve depends on the temporal sampling as well as the intrinsic variability of the source. The sampling frequency in the J-band is better than that in the B-band. Hence, we resampled the $\gamma$-ray and J-band light curves with that in the B-band. To carry out the resampling, we remove data points from the $\gamma$-ray and J-band light curve segments for which there are no corresponding B-band data point within $\pm0.5$ days. This was done to ensure that the comparison of the ACFs is physically meaningful. In 3 out of 6 blazars in our sample, namely, AO 0235+164, PKS 1510-089, and 3C 454.3, the B-band as well as the resampled $\gamma$-ray and J-band light curves had a sampling of at least once in 3 days. The ACFs of those 3 sources at all three wave bands are shown in Figure \ref{acf1}. For the ACF calculation, we use a bin-width of 5 days. We list the half-width at half maximum (HWHM) of the central peaks of the ACFs in Table \ref{acf_table}. The ACFs in all bands are very similar in shape for AO 0235+164 and 3C 454.3. In PKS 1510-089, the $\gamma$-ray ACF is slightly narrower in shape but the HWHM is very similar to the that in B and J-bands. Since the temporal sampling in all 3 bands are the same, this is a comparison of the intrinsic variability properties of these three blazars. This shows that the characteristic size-scales of the corresponding emission regions in all three bands are similar which, in turn, may imply that the $\gamma$-ray and optical-IR emission in these blazars is generated by electrons of similar energies. Even initially homogeneous electron acceleration region will develop energy-dependent stratification due to the energy dependence of radiative loss timescales, i.e., lower energy electrons naturally diffuse into a larger volume. Similarly, a higher energy population of electrons will be in a smaller volume causing the ACF width and characteristic timescales of variability to be smaller.
\begin{figure*}
\includegraphics[height=7in,width=2in,angle=-90]{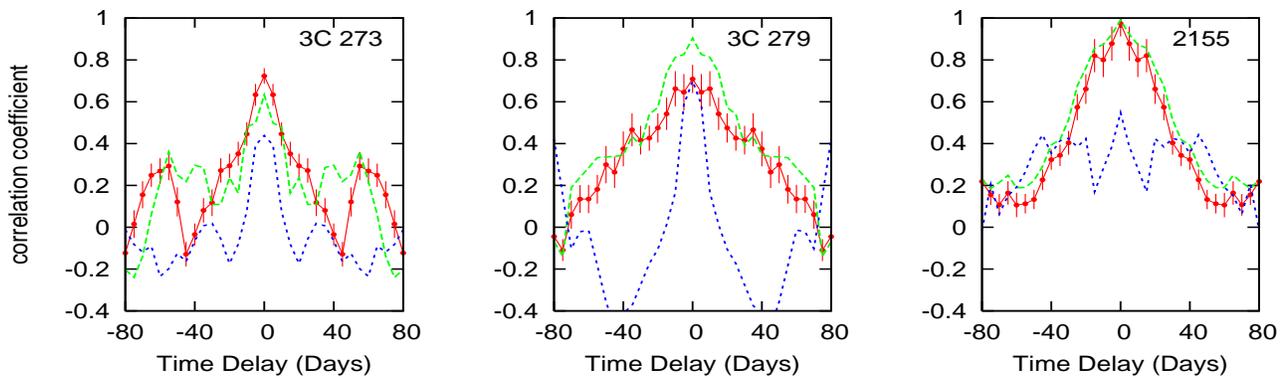}
\caption{Discrete auto-correlation functions of the $\gamma$-ray (blue dotted line), $B$-band (red solid line and filled circles with error bars), and $J$-band (green dashed line) light curves of 3C 273, 3C 279, and PKS 2155-304. Because of the scarcity of data, sampling differed at optical/IR and $\gamma$-ray wavelengths. Uncertainties shown by the red data points are characteristic of other bands as well.}
\label{acf2}
\end{figure*}

Figure \ref{acf2} presents the ACFs of the other 3 blazars, namely, 3C 273, 3C 279 and PKS 2155-304, for which resampled data were not used due to poor sampling. In 3C 273, the ACFs at different bands are similar while in 3C 279, the $\gamma$-ray ACF is significantly narrower. This may be an effect of the temporal sampling and can not be confirmed with the present data. The peak of the $\gamma$-ray ACF of the blazar 2155-304 is much smaller than the rest. This is because, the $\gamma$-ray variation of this source was dominated by short-term flares uncorrelated to each other. This behavior is also reflected in the power spectrum of its variation as discussed in {\S}3. We do not find any significant peak in any of the ACFs other than the central peak. That implies the absence of significant quasi-periodic behavior in the variability of these blazars in $\gamma$-ray and optical-IR frequencies.

\section{Power Spectral Analysis}
We use a variant of the Power Spectrum Response method \citep[PSRESP;][]{utt02} to determine the intrinsic PSD of the optical light curves. Our realization of PSRESP is described in \citet{cha08} in details. Here we briefly summarize the method.
\begin{figure*}
\includegraphics[height=7in,width=5in,angle=-90]{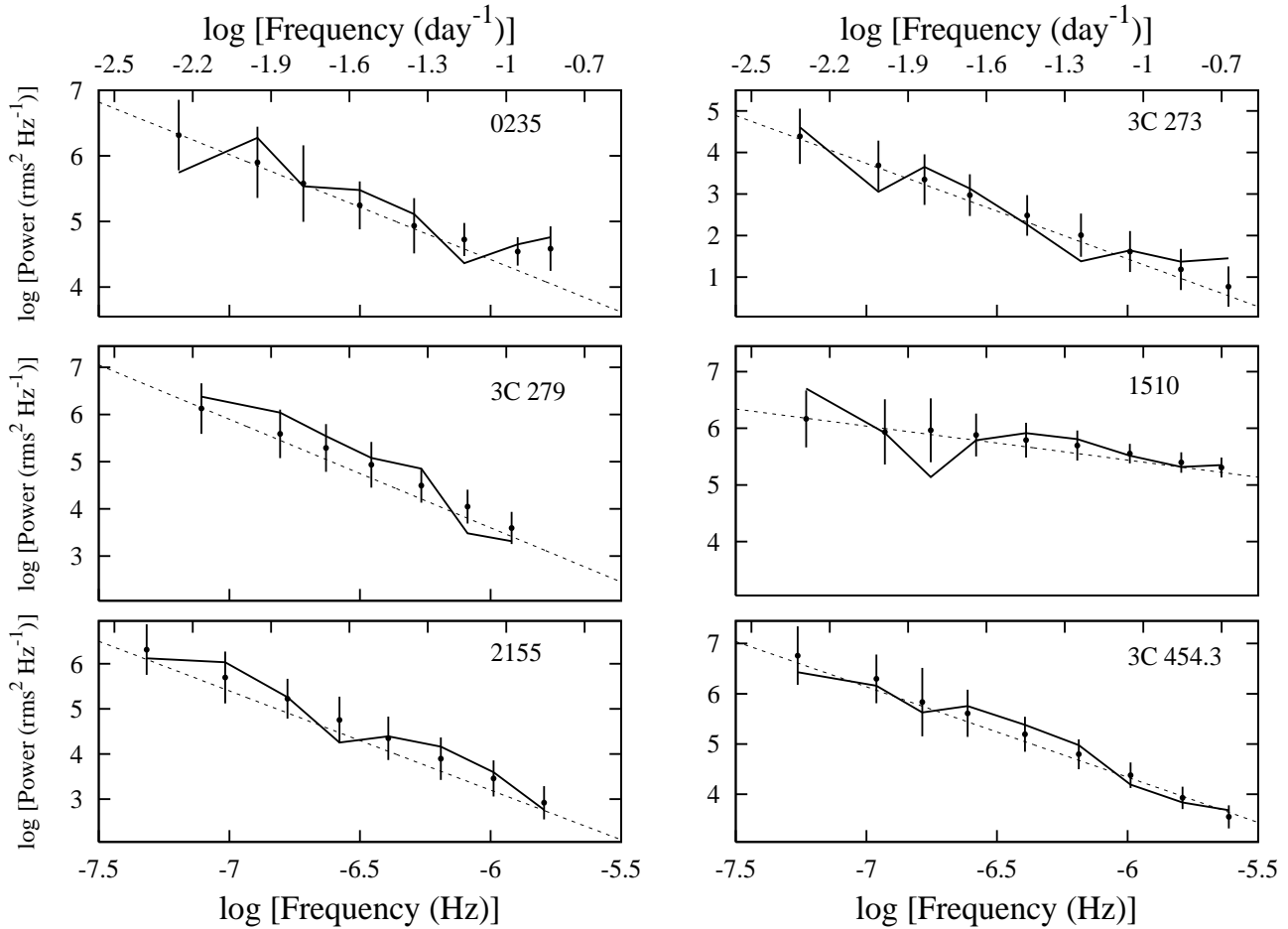}
\caption{Power spectral density of the $R$-band light curves of the six blazars in our sample (calculated using the PSRESP method). The PSD of the observed data is given by the solid jagged line, while the underlying power-law model is given by the dashed straight line. Filled circles with error bars correspond to the mean value of the PSD simulated from the underlying power-law model. The error bars are one standard deviation of the distribution of simulated PSDs. The slopes of the best-fit models are listed in Table \ref{psd_table}. All PSDs are well-fit by simple power-law models with no break.}
\label{psdall}
\end{figure*}

We start with an underlying model for the PSD function, such as a simple power-law. Then we simulate $N$ light curves starting from this underlying model using the technique described in \citet{tim95}. We use $N=100$. We resample these simulated light curves with the sampling window of the observed light curve. We also add other noise to the simulated light curves which affect the real observations due to its finite length (``red noise leak") and discontinuous sampling (``aliasing"). Then we calculate the PSD of the real light curve (denoted by PSD$_{\rm obs}$) and that of each of the simulated light curves (denoted by PSD$_{\rm sim,i}$, i=1, N). We calculate the approximate value of the power due to Poisson noise using the observational uncertainties and add that power to PSD$_{\rm sim,i}$. We finally compare PSD$_{\rm obs}$ with the mean of the PSD$_{\rm sim,i}$ (denoted by $\overline{\rm PSD}_{\rm sim}$) weighted by the standard deviation in the distribution PSD$_{\rm sim,i}$ to determine the goodness of fit of the underlying model that we assumed at the beginning. The goodness of fit is quantified by the ``success fraction'' $F_{\rm succ}$ defined as the fraction of simulated light curves that successfully represent the observed light curve. We do this for a range of underlying models covering a large parameter space and determine the underlying model that provides the best-fit, i.e., the highest $F_{\rm succ}$. The observed PSD suffers from the distorting effects of the finite length (``red noise leak") and discontinuous sampling (``aliasing") of the light curves as well as power generated by the temporal sampling. These are accounted for by the method that we use.

Similar to the ACF calculation, we select the longest segments of SMARTS R-band light curves of the six blazars without seasonal gaps ($\sim$3-4 months). This is done in order to minimize the distortion of the power spectrum and to cover the largest possible range of timescales. The length of such segments were 200$-$250 days for all six blazars in our sample and the average sampling rate was one data point every 2-3 days. We chose the R-band for this calculation because among the optical bands it has the best sampling for all objects.
\begin{figure}
\epsscale{.80}
\plotone{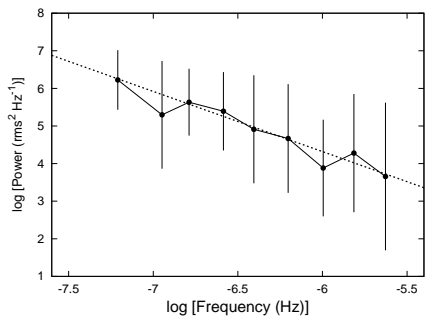}
\caption{The data points show the average power spectral density of the six blazars in our sample. The error bars are one standard deviation about the mean. The average slope $-1.6\pm0.2$ is denoted by the dotted line.}
\label{avgpsd}
\end{figure}

Figure \ref{psdall} presents the PSDs of the six blazars and the corresponding power-law fits. The solid jagged line shown in Figure \ref{psdall} is PSD$_{\rm obs}$ and the filled circles with error bars are the mean ($\overline{\rm PSD}_{\rm sim}$) and standard deviation of the distribution PSD$_{\rm sim,i}$. The resampling and addition of the noise described above add extra power to $\overline{\rm PSD}_{\rm sim}$ causing the filled circles with error bars to be slightly higher than the underlying model (dashed line) at certain frequencies. 

The optical R-band PSD of all blazars in our sample show red noise behavior, i.e., there is higher amplitude variability on longer than on shorter timescales. We list the best-fit slope and success fraction of the PSD of all blazars in Table \ref{psd_table}. The rejection confidence, equal to one minus the success fraction, is much less than 0.9 in all cases. This implies that a simple power-law model provides an acceptable fit to all the PSDs. We also fit a broken power-law model to all the PSDs, setting the low-frequency slope at $-1.0$ and allowing the break frequency and the slope above the break over a wide range of parameters ($10^{-7}$ to $10^{-6}$ Hz and $-1.0$ to $-2.5$, respectively) while calculating the success fractions. This gives lower success fractions than the simple power-law model across the entire parameter space. Hence, there is no significant break detected in the PSD of any of the sources.  We determine the optical PSD up to the highest variational frequency that can be achieved with the existing data. A better constraint on the existence of a break in the optical PSD might be achieved with a broader range of sampled frequencies. 

Among the blazars in our sample, the $R$-band variability amplitude of 3C 273 was much smaller than the others during the time interval considered here. This is evident in the PSD which shows that the power of variability in 3C 273 is about two orders of magnitude less than the average of the other blazars at all timescales. The PSD of the blazar PKS 1510-089 has a significantly flatter slope than the other blazars in our sample indicating larger amplitude of short-timescale variations. This trend is evident in the light curve of this blazar which is dominated by very large-amplitude flares of width a few days.
\begin{table}
\begin{center}
\caption{Slope and Success Fraction of the Best-Fit Power Spectral Models.\label{psd_table}}
\begin{tabular}{ccc}\\
\tableline\tableline
Object & Slope & $F_{succ}$ \\
\tableline
AO 0235+164		&	$-1.6^{+0.3}_{-0.3}$	&	0.74	\\
3C 273			&	$-2.3^{+0.2}_{-0.5}$	&	0.63	\\
3C 279			&	$-2.3^{+0.5}_{-0.2}$	&	0.99	\\
PKS 1510-089		&	$-0.6^{+0.5}_{-0.2}$	&	0.87	\\
PKS 2155-304		&	$-2.2^{+0.2}_{-0.4}$	&	0.93	\\
3C 454.3		&	$-1.8^{+0.3}_{-0.3}$	&	0.81	\\	
\tableline 
\end{tabular}
\end{center}
\end{table}

The average power spectral density of the six blazars in our sample is shown in Figure \ref{avgpsd}. The data points and the error bars are the mean and standard deviation of the power of all six blazars in each logarithmic bin. The best fit slope of the average PSD is $-1.6\pm0.3$, where the uncertainty is one standard deviation about the mean. \citet{abd10_timing} found that the average slope of the best-fit power-law for a sample of 9 bright FSRQs is $-1.4\pm0.1$ and that of 6 bright BL Lacs is $-1.7\pm0.3$, similar to the slope we found. This is consistent with the model where the emission at $\gamma$-ray energies and optical frequencies are both generated by electrons through synchrotron and inverse-Compton processes, respectively, so that the nature of variability at these two bands is similar. The uncertainty in the individual as well as the average PSD will decrease when more data are included and may reveal the existence of small difference in the amplitude of power between the $\gamma$-ray and optical-IR variability at a given frequency. For example, the optical PSD of the BL Lac object PKS 2155-304 is quite steep (slope $\sim -2.2$) while the $\gamma$-ray PSD is much flatter \citep[see Fig. 17 of][]{abd10_timing}. This implies that the short-term (few to 10 days) variability in this blazar is more prominent at $\gamma$-ray energies than at optical wave bands. Possible explanation for this is discussed in {\S}6.

\section{Shortest Timescales of Variability}
We calculate the doubling timescales ($T_{\rm doub}$) for each object in all bands. $T_{\rm doub}$ is defined as the minimum time interval over which the flux increases by a factor of 2. We scan the light curves of all objects in our sample in each band to find time intervals over which the flux has doubled and then select the minimum of those time intervals as the doubling timescale for the respective band. Similarly, we also determine the minimum timescale over which the flux has decreased by a factor of 2 and denote it by $T_{\rm half}$. To avoid spurious results, we select $T_{\rm doub}$ and $T_{\rm half}$ only in those cases where increase or decrease of flux at that timescale has happened 3 or more times during the period we scanned and the general trend is present at all bands. 
\begin{deluxetable*}{cccccccccccccc}
\tablecolumns{14} 
\tablewidth{0pt}
\tablecaption{Variability Timescales Present in the Multi-Frequency Light Curves of the Six Blazars. For the sources AO 0235+164 and PKS 1510-089, and the $\gamma$-ray band, all timescales are upper limits.\label{doubling_table}}
\tablehead{
	&	\multicolumn{6}{c}{T$_{\rm double}$ (days)} &  \multicolumn{6}{c}{T$_{\rm half}$ (days)} \\
\cline{2-7} \cline{9-14} \\ 
\colhead{Object} 	& \colhead{$B$}   & \colhead{$V$} & \colhead{$R$}   & \colhead{$J$}   & \colhead{$K$}	&\colhead{$\gamma$-ray}	&\colhead{}   & \colhead{$B$} & \colhead{$V$}  & \colhead{$R$}   & \colhead{$J$}   & \colhead{$K$}	&\colhead{$\gamma$-ray}}
\startdata
AO 0235+164 & $<$1.9    & $<$1.9 & $<$1.9    & $<$1.9   & - 	& $<$1.0 && $<$2.0   & $<$2.0   & $<$2.0 & $<$2.0  & - &$<$1.0   	\\
3C 273			& -       & -     & -       & -       & - 	&$<$1.0&      & -    & -       & -       & -      & -  		&$<$1.0  	\\
3C 279			& 4.1   & 4.1  & 4.2   & 4.2    & 4.1	&$<$1.0&      & 4.9 & 4.9   & 4.9    & 4.9   & 4.9  	&$<$1.0 	\\
PKS 1510-089& $<$2.0   & $<$2.0  & $<$2.0   & $<$2.0   & $<$2.0	&$<$1.0&      & $<$1.9  & $<$1.9   & $<$1.9    & $<$1.9   & $<$1.9 	&$<$1.0 	\\
PKS 2155-304	& 32.0 & 36.0 & 32.0 & 33.0 & 30.0	&$<$1.0&      & 23.0 & 22.0 & 25.0  & 23.0 & 25.0  	&$<$1.0 	\\
3C 454.3			& 3.9   & 3.9  & 2.9   & 2.9   & 1.9	&$<$1.0&      & 10.0 & 9.1   & 12.0  & 9.1   & 10.0  	&$<$1.0 	
\enddata
\end{deluxetable*}

Table \ref{doubling_table} lists $T_{\rm doub}$ and $T_{\rm half}$ for all objects in our sample in all available bands. It can be seen from the table that the timescales over which the flux increases or decreases by a factor of 2 is similar in a given band for an object. This broad similarity implies that the flares are symmetric and hence crossing time dominated. However, the value of $T_{\rm half}$ is larger than $T_{\rm doub}$ in 3C 454.3. This blazar was not observed by SMARTS during its decay from the large outburst in 2009 December due to Sun-gap. The faster decay and hence a smaller value of $T_{\rm half}$ could be detected during that time.

\begin{figure}
\epsscale{.80}
\plotone{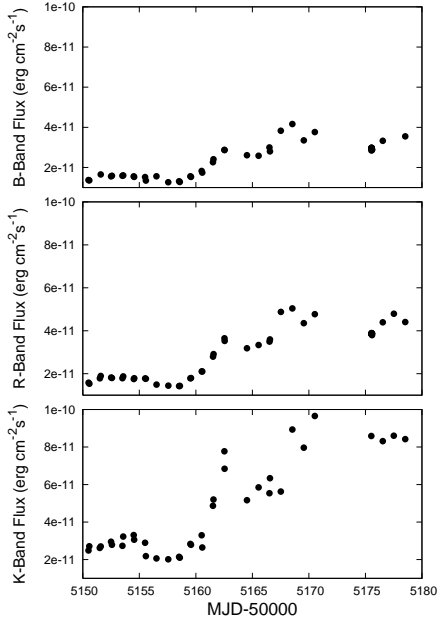}
\caption{Variation of 3C 454.3 at $B$, $R$, and $K$ band during the large outburst in 2009 December. It is clear that the variability amplitude is larger at the longer wavelengths. This may be due to the ``dilution" of the variability due to the presence of a bluer emission from an accretion disk-like component which stays constant during this interval.}
\label{shortterm1}
\end{figure}

No object shows a significant difference in those timescales among the optical-near IR bands. However, in 3C 454.3, the doubling timescale is smaller at longer wavelength optical-IR bands. This may be due to the disk contribution in the bluer bands \citep[e.g.,][]{bonnoli11} which dilutes the jet variability and hence increases the doubling timescale. This is supported by observation of 3C 454.3 at B, V, R, and J bands during 2009 December as shown in Figure \ref{shortterm1}. The B-band flux changes by a factor of only $\sim$1.5 over 2--4 days while the corresponding change in J-band is by a factor of $\sim$6. 

$T_{\rm doub}$ and $T_{\rm half}$ of all 6 objects in the $\gamma$-ray band is $\sim$1.0 day. These are all upper limits since the timescales are determined from light curves provided by the Fermi team which are binned at 1-day intervals. Those timescales for AO 0235+164 and PKS 1510-089 are $\sim$2 days in all optical-IR bands, consistent with those in the $\gamma$-ray band. These are also upper limits since we do not have data points between those 2 days to check if the actual timescales are significantly less than that. In the case of 3C 273, the flux value does not change by a factor of 2 over any timescale during the two yr of monitoring. In 3C 279, those timescales at all optical-IR bands is $\sim$4-5 days. This is not an upper limit since we do have data points during those 4 days and the flux did not change by factor of 2 in less than 4 days. The slower variability indicated by the longer doubling timescales in optical-IR bands than that in the $\gamma$-rays is consistent with the narrower auto-correlation function of 3C 279 in the latter. This may imply that in 3C 279, the inverse-Compton (IC) $\gamma$-ray emission is generated by a slightly higher energy population of electrons than the synchrotron optical emission. In PKS 2155-304, the smallest timescales over which the optical-IR flux changed by a factor of 2 is $\sim$30 days, again suggesting slower variability in these wave bands similar to the steeper power spectral slope in optical-IR than that in the $\gamma$-rays as found in {\S}3. This is discussed in more details in {\S}6.
\begin{figure*}
\includegraphics[height=6in,width=5in,angle=-90]{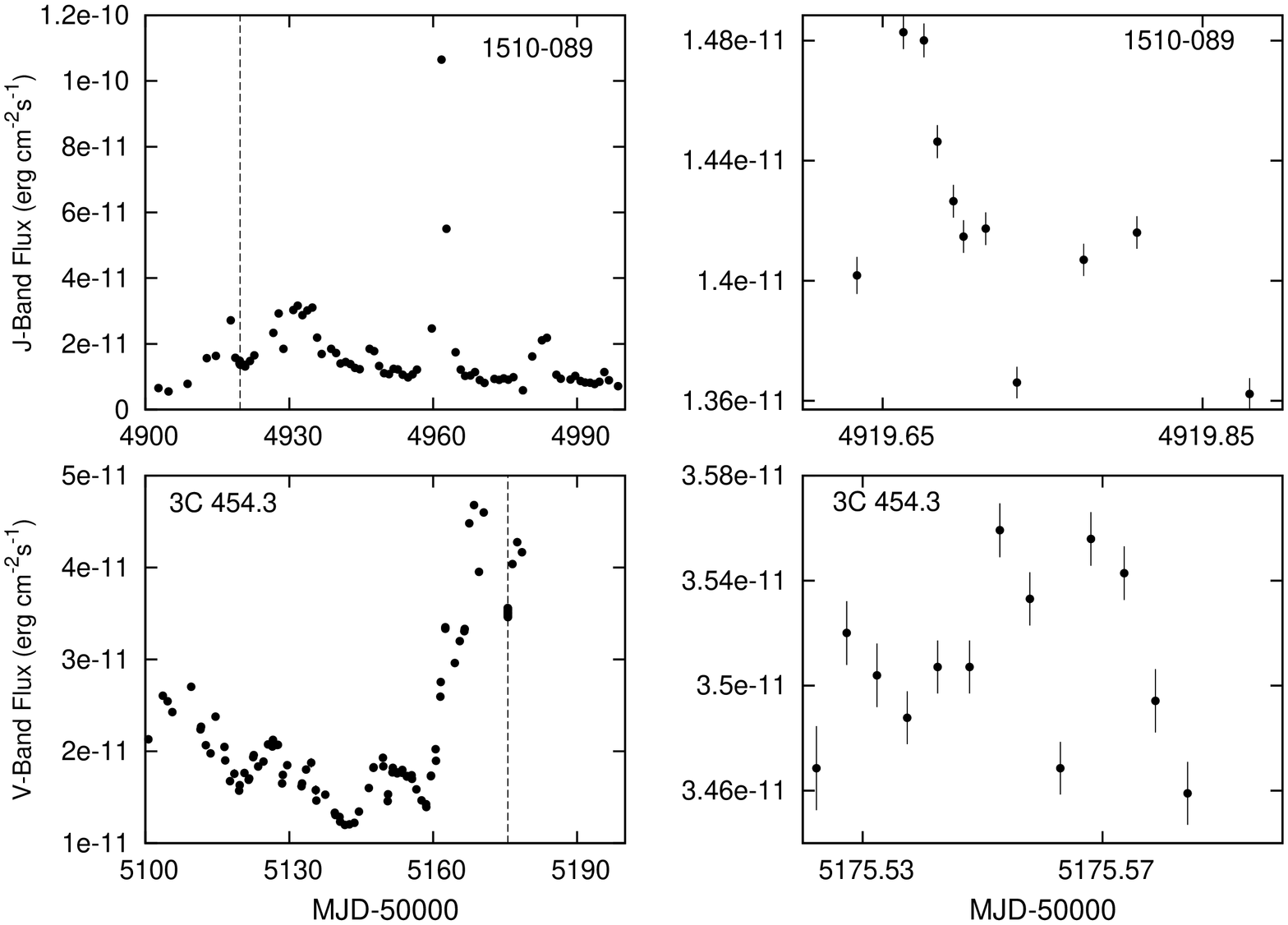}
\caption{The right panels show the very short-term (few hours) variability of the blazars PKS 1510-089 (J-band) and 3C 454.3 (V-band). The corresponding panels on the left show the longer-term variability of the same object at the same wave band. The dashed vertical lines in the left panels denote the intervals shown in the right panel. There is significant variability at timescales of a few hours in both objects but the amplitude is $\sim 5-10$\%, much smaller than the $\gamma$-ray variability at these timescales.}
\label{shortterm2}
\end{figure*}

\citet{tav10_location} showed that during a very high state in 2009 December, the $\gamma$-ray flux from 3C 454.3 changed by a factor of $\sim$5 at 6--12 hr timescale. Our observations of this object during that interval, as shown in Figure \ref{shortterm1}, had a sampling rate of only $1-2$ times per night. We have obtained observations of 3C 454.3 and PKS 1510-089 to search for intra-night variability during other nights. We show these light curves in Figure \ref{shortterm2}. They show significant variability over the few hour timescales but the amplitude of variability ($\sim$5-10\%) is not at the same level as the hour-scale $\gamma$-ray variability of 3C 454.3 and PKS 1510-089 shown in \citet{tav10_location}. The intra-night optical-IR variability shown here is not simultaneous with the very high state of $\gamma$-ray emission when $\gamma$-ray light curves with comparable time resolution can be extracted with enough signal to noise ratio. Stronger constraints on the relation between $\gamma$-ray/optical-IR emission could be drawn from such data if they were exactly simultaneous.

\section{Flare Analysis}
Blazar light curves at all wave bands have been interpreted as a superposition of outbursts caused by events in the jet or the accretion disk/corona region in addition to a steady baseline flux \citep[e.g.,][]{val99,cha08,jor10,abd10_timing}. Therefore, comparison of the properties of individual contemporaneous flares at the $\gamma$-ray and optical-IR wavebands is a potential diagnostic of their origin. To investigate this, we follow \citet{val99} to decompose the light curves into individual (sometimes overlapping) flares, each with exponential rise and decay of the form:\\
\begin{equation}
{\rm f(t)=f_0 + f_{\rm max}exp[(t-t_0)/T_r], ~~~~for ~t<t_0, and}
\end{equation}
\begin{displaymath}
{\rm     ~~=f_0 + f_{\rm max}exp[-(t-t_0)/T_d] ~~~~ for ~t>t_0}
\end{displaymath}
\begin{figure*}
\epsscale{0.8}
\plotone{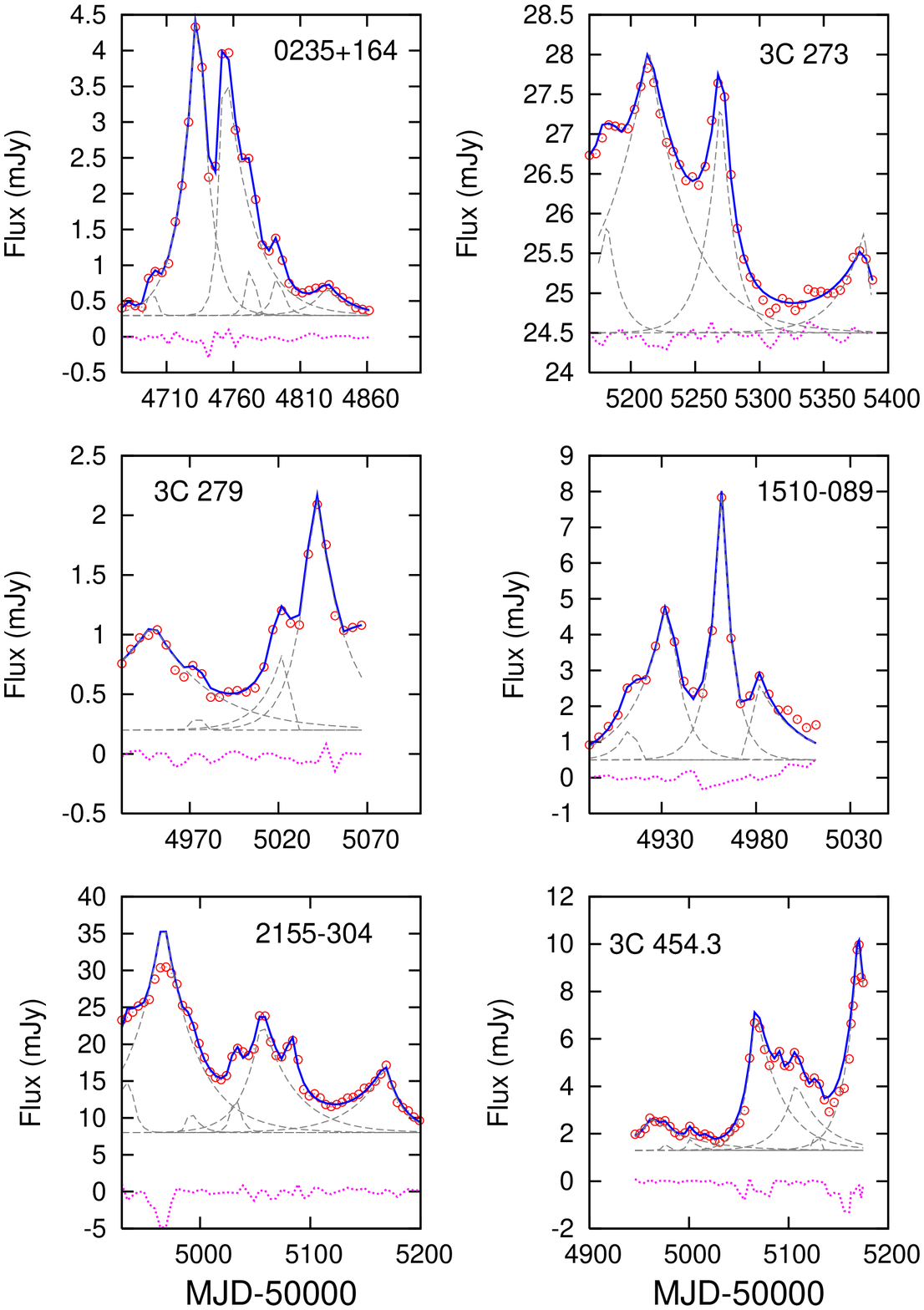}
\caption{Decomposition of continuous optical light curves of six blazars into individual flares. Red points denote the smoothed R-band light curves of AO 0235+164, 3C 273, 3C 279, PKS 1510-089, PKS 2155-304, and 3C 454.3. Blue solid curves correspond to summed flux after modeling the light curve as a superposition of several individual flares and a constant background level, magenta dotted lines show the residual fluxes in each case, and gray dashed lines denote the underlying individual flares.\\}
\label{opfit}
\end{figure*}
In this equation, f$_0$ is the background level of flux that stays constant over the corresponding interval, ${\rm f_{max}}$ is the amplitude of the flare, ${\rm t_0}$ is the epoch of the peak, and ${\rm T_r}$ and ${\rm T_d}$ are the rise and decay time scales, respectively. These are the five free parameters for each flare. $f_0$ was constrained to be less than or equal to the lowest value of the flux during the respective interval. There was no prior restriction on the possible values of the other four free parameters. Exponential rise and decay is a widely used and successful model for observationally well-sampled flux outbursts in blazars, at all wavelengths \citep[e.g.,][]{val99,abd10_timing}. We have tried other models such as linear rise and decay and a Gaussian profile but the fit is better with the exponential profile described in the paper.

Similar to {\S}2 and {\S}3, we select the longest segments of R-band and $\gamma$-ray light curves of the six blazars without seasonal gaps ($\sim$3-4 months) to increase the accuracy of the flare decomposition analysis. Before the decomposition, we smooth the light curve using a Gaussian function with a 5-day FWHM smoothing time. Our goal is to compare the properties of the longer-term (longer than 5 days) flares present in the optical light curves. We have shown that the PSDs of the optical variability correspond to red noise, i.e., there is higher amplitude variability on longer than on shorter timescales. Because of this, we analyze the more powerful longer timescale flares than the relatively weak flares on small time scales. The nature of the following results does not change if we use a 3-day or a 7-day FWHM for the smoothing function.
\begin{figure*}
\epsscale{.80}
\plotone{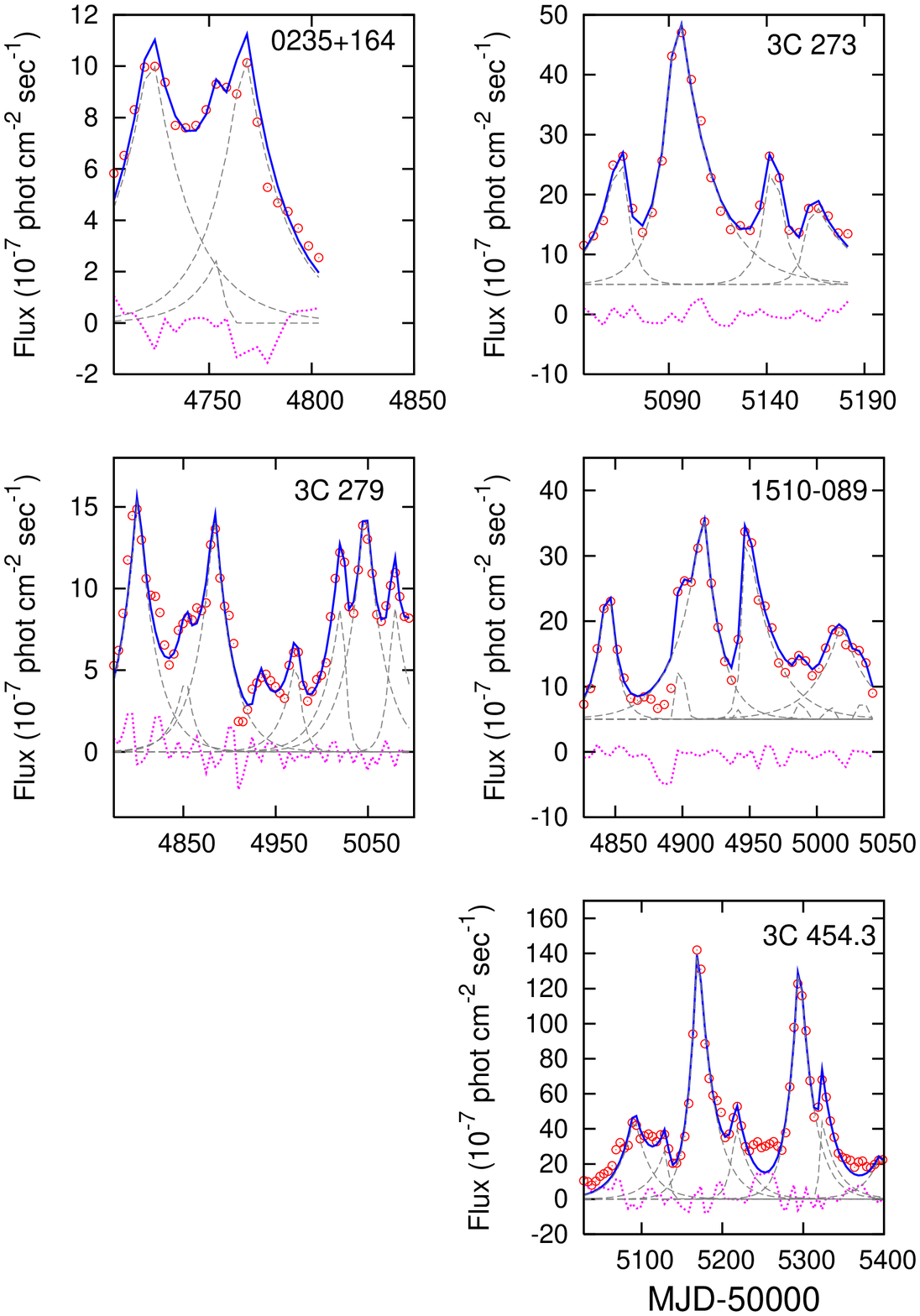}
\caption{Decomposition of continuous $\gamma$-ray light curves of six blazars into individual flares. Red points denote the smoothed 0.1-300 GeV light curves of AO 0235+164, 3C 273, 3C 279, PKS 1510-089, and 3C 454.3. Blue solid curves correspond to summed flux after modeling the light curve as a superposition of several individual flares and a constant background level, magenta dotted lines show the residual fluxes in each case, and gray dashed lines denote the underlying individual flares. Similar analysis for PKS 2155-304 was not carried out since in that source the variability is dominated by sporadic short-term flares with frequent non-detection at intermediate times which is not suitable for fitting with the function that we use.}
\label{gammafit}
\end{figure*} 
\begin{deluxetable}{ccccccc}
\tablecolumns{7}
\tablewidth{0pt}
\tablecaption{The best-fit parameters of the major individual flares in R-band.\label{flareparam_op}}
\tablehead{
\colhead{} &\colhead{} &\multicolumn{5}{c}{Flare Parameters}\\
\cline{3-7}\\
\colhead{Object} & \colhead{Flare Index} & \colhead{${\rm f_{max}}$\tablenotemark{a}} & \colhead{${\rm t_0}$\tablenotemark{b}} & \colhead{${\rm T_r}$\tablenotemark{c}} & \colhead{${\rm T_d}$\tablenotemark{c}} & \colhead{$\xi$\tablenotemark{d}}}
\startdata
AO 0235+164	& op1	& 4.5 	&  4734	&  12.5	&  9.0	& -0.2	\\
		& op2	& 3.9	&  4753	&  4.0	&  20.0	&  0.7	\\
3C 273		& op1 	& 28.0	&  5214	&  40.0	&  40.0	&  0.0	\\
		& op2	& 27.6 	&  5270	&  11.0	&  8.5	& -0.1	\\
3C 279		& op1	& 2.1	&  5040	&  10.0	&  17.0	&  0.3	\\
		& op2	& 1.1 	&  4949	&  39.5	&  31.5	& -0.1	\\
PKS 1510-089	& op1	& 7.8 	&  4962	&  7.0	&  6.5	&  0.0	\\
		& op2	& 4.8 	&  4934	&  16.5	&  7.5	& -0.4	\\
PKS 2155-304	& op1	& 36.0 	&  4966	&  30.0	&  30.0	&  0.0	\\
		& op2	& 22.0 	&  5056	&  17.0	&  30.0	&  0.3	\\
3C 454.3	& op1	& 10.2 	&  5170	&  13.5	&  19.5	&  0.2	\\
		& op2	&  7.3	&  5067	&  11.0	&  26.5	&  0.4	\\
\enddata
\tablenotetext{a}{Flare amplitude in units of mJy.}
\tablenotetext{b}{Date of peak in units of MJD-50000.}
\tablenotetext{c}{Rise and decay timescales in units of days.}
\tablenotetext{d}{Skewness parameter: $\xi=\frac{T_d-T_r}{T_d+T_r}.$}
\end{deluxetable}
\begin{deluxetable}{ccccccc}
\tablecolumns{7}
\tablewidth{0pt}
\tablecaption{The best-fit parameters of the major individual flares in $\gamma$-rays.\label{flareparam_gam}}
\tablehead{
\colhead{} &\colhead{} &\multicolumn{5}{c}{Flare Parameters}\\
\cline{3-7}\\
\colhead{Object} &\colhead{Flare Index}&\colhead{${\rm f_{max}}$\tablenotemark{a}} & \colhead{${\rm t_0}$\tablenotemark{b}} & \colhead{${\rm T_r}$\tablenotemark{c}} & \colhead{${\rm T_d\tablenotemark{c}}$} & \colhead{$\xi$\tablenotemark{d}}}
\startdata
AO 0235+164	& g1 &	11.0	&  4767  &  16.5	&  20.0	&   0.1	\\
		& g2 &	11.0	&  4722  &  20.0	&  20.0	&   0.0	\\
3C 273		& g1 &	49.0	&  5094  &   9.0	&  17.5	&   0.3	\\
		& g2 &	25.5	&  5066  &  12.5	&  5.0	&  -0.4	\\
		& g3 &	24.5	&  5144  &   6.0	&  6.0	&   0.0	\\
3C 279		& g1 &	16.0	&  4801  &  20.0	&  20.0	&   0.0	\\
		& g2 &	15.8	&  5047  &  20.0	&  20.0	&   0.0	\\
		& g3 &	14.8	&  4884  &  20.0	&  19.5	&   0.0	\\
		& g4 &	11.0	&  5023  &  15.0	&  4.0	&  -0.6	\\
		& g5 &	9.7	&  5078  &  7.0		&  20.0	&   0.5	\\
PKS 1510-089	& g1 &	35.0	&  4916  &  19.0	&  12.5	&  -0.2	\\
		& g2 &	34.0	&  4947  &  3.5		&  20.0	&   0.7	\\
		& g3 &	24.0	&  4845  &  8.5		&  7.0	&  -0.1	\\
		& g4 &	19.0	&  5019  &  20.0	&  20.0	&   0.0	\\
3C 454.3	& g1 &	140.0	&  5170  &  12.0	&  20.0	&   0.3	\\
		& g2 &	130.0	&  5295  &  13.5	&  18.0 &   0.1	\\
\enddata
\tablenotetext{a}{Flare amplitude in units of $10^{-7}$ ph cm$^{-2}$s$^{-1}$.}
\tablenotetext{b}{Date of peak in units of MJD-50000.}
\tablenotetext{c}{Rise and decay timescales in units of days.}
\tablenotetext{d}{Skewness parameter: $\xi=\frac{T_d-T_r}{T_d+T_r}.$}
\end{deluxetable}

We proceed by first fitting the highest peak in the smoothed light curve to an exponential rise and decay, and then subtracting the flare thus fit from the light curve. We do the same to the ``reduced" light curve, i.e., we fit the next highest peak. This reduces confusion created by a flare already rising before the decay of the previous flare is complete. We fit the entire light curve in this manner with a number of individual (sometimes overlapping) flares, leaving a residual flux much lower than the original flux at all epochs. In each case, we use the minimum number of flares required to adequately model the light curves such that using more flares does not change the residual flux by more than 10\%. Figures~\ref{opfit} and \ref{gammafit} compare the smoothed optical and $\gamma$-ray light curves with the summed flux (sum of contributions from all the model flares at all epochs). The $\gamma$-ray light curve of the blazar PKS 2155-304 was not modeled in the above analysis. That is because the Fermi $\gamma$-ray light curve of this blazar is dominated by sporadic short-term flares with frequent non-detection at intermediate times which is not suitable for fitting with the function that we use. Among all the flares used to adequately fit the light curves, we show the best-fit parameters of the major flares in Tables \ref{flareparam_op} and \ref{flareparam_gam}. We define the major flares as the ones whose amplitudes are at least at the 50\% level of the highest peak of that light curve. We consider the the major flares only because in the case of the smaller flares the rise and decay timescales are not well-constrained. 

\begin{figure}
\epsscale{.80}
\plotone{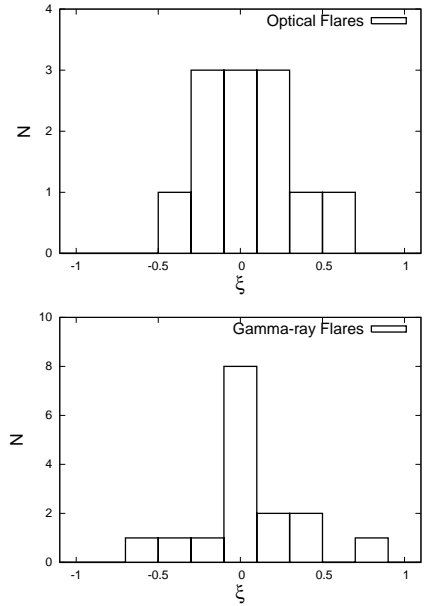}
\caption{Distribution of the skewness parameter ($\xi$) for the major optical and $\gamma$-ray flares. This shows that most of the large flares are symmetric in nature ($|\xi| < 0.3$) at both wave bands. }
\label{skewness}
\end{figure}

We define a skewness parameter ($\xi$) as the following:\\
\begin{equation}
\xi=\frac{T_d-T_r}{T_d+T_r}.
\end{equation}
$\xi$ is $0$ for exactly symmetric flares. If the decay is slower than the rise then $\xi$ is positive and \textit{vice versa}. We show the distribution of the skewness parameter for all major optical and $\gamma$-ray flares in Figure \ref{skewness}. This indicates that in both wave bands most of the major flares are symmetric ($|\xi| < 0.3$). The rise and decay times were two separate free parameters with no constraints. Hence, the best-fit models were free to contain widely asymmetric values of the rise and decay timescales. We showed that the majority of the large and well-defined flares were roughly symmetric. This result is not due to any constrain in our modeling process but comes out naturally from the data. 

The variability in the emission from blazar jets is primarily caused by the interaction of a shock wave with the jet plasma \citep[e.g.,][]{bla79,mar85} over a finite size or a finite time interval. The former can occur when a disturbance in the jet-flow or a blob of denser plasma, observed as moving knots in radio-interferometric observations, passes through a standing shock present in the jet. The latter can occur due to the evolution of an internal shock. Due to the existence of a finite size and the resultant finite duration of the above events, the observations are affected by light-travel time. Any process faster than the light-travel time through the said size or the finite duration of the event ($\Delta$t) will be smoothed out \citep[e.g.,][]{chi99,kat00,sok04,che11}. If the radiative cooling time is longer than $\Delta$t, then we would expect to see a significantly longer decay time than rise time. This is because the radiative cooling timescale of electrons which are generating the optical-IR emission through synchrotron radiation and $\gamma$-rays through IC processes is longer than time needed to inject particles (usually assumed to be ``instantaneous") responsible for the flare into the region where they would be able to radiate significantly. On the other hand, we may expect to see symmetric flares, i.e., the rise and decay times are comparable, if the radiative cooling time of the emitting particles is much smaller than $\Delta$t. For example, in the case of a disturbance in the jet moving through a standing shock, the rise time corresponds to the time it takes for the entire disturbance to enter the standing shock region while the number of energized electrons increases resulting in an increase in flux. Similarly, the decay time corresponds to the time taken by the disturbance to leave the standing shock region entirely while the number of energized electrons steadily decreases since the radiative cooling times are much smaller than the rise and decay times. In this case, the rise and decay times will be similar resulting in a symmetric flare. The symmetric nature of the observed flares in this work indicates that the rise and decay timescales are dominated by crossing time of radiation or a disturbance through the emission region, similar to the latter case described above.

\section{Difference in the Time Variability Properties of FSRQs and BL Lacs}
The main difference between the time variability properties of these two classes of objects that has been revealed in this work is that the R-band variation is significantly smoother than that in the $\gamma$-ray energies in the only BL Lac object in our sample, namely, PKS 2155-304 while in the FSRQs the variations in these two wave bands are similar. This may be caused by one or both of the following reasons:\\
(1) The $\gamma$-ray emission in blazars is generated by the IC processes. Hence, the variability is a combination of the variation of the emitting electrons and that of the seed photons which are being inverse-Compton scattered to the $\gamma$-ray energies. In the case of the HBLs such as PKS 2155-304, GeV $\gamma$-ray emission is dominated by the SSC process while in the FSRQs it is dominated by the EC contribution. In the former, the seed photons which are produced by the synchrotron emission in the jet are varying faster than those in the latter which are generated in the disk, broad line region or the torus. Hence, in the case of PKS 2155-304, combined variability of the emitting electrons and the seed photons which is reflected in the GeV band is faster than that in the R-band which represents the variability of the synchrotron emission only. This is not seen in the FSRQs where little or none of the GeV variability is due to the variation in the slowly varying seed photons. Hence the GeV and R-band variability is very similar because both are contributed by the variation in the emitting electrons only.\\
(2) From the SED of PKS 2155-304 \citep{abd10_sed}, it can be seen that the R-band emission is $\sim$1.5 orders of magnitude lower than the synchrotron peak while the $\gamma$-ray peak is at 4 GeV, which is within the wave band we are considering in this paper, 0.1-300 GeV. The photon index of this object from the \textit{Fermi} 2-yr catalog \citep{abd11_2yrcatalog} is 1.84. This means in this blazar, the 0.1-300 GeV emission has significant contribution from electrons which are at a higher energy than those producing the synchrotron radiation unlike the FSRQs where the GeV and R-band emission is produced by electrons of similar energies. This may partially contribute to the faster variability in the GeV band than that in the R-band in PKS 2155-304, as observed.

\section{Detailed Comparison of the $\gamma$-ray and Optical Light Curves During a Prominent Multi-frequency Outburst of the Blazar 3C 454.3}
The flat spectrum radio quasar 3C 454.3 went through a very large multi-wavelength outburst during 2009 December. There were several smaller but significant flares at $\gamma$-ray energies as well as optical-IR frequencies during the four months prior to the large outburst. To investigate the location of these outbursts at $\gamma$-ray energies as well as to compare their properties with corresponding flares at optical-IR frequencies, if present, we decompose the light curves of 3C 454.3 from Fermi and SMARTS into individual, sometime overlapping flares in the same manner as in {\S}5.

\begin{figure}
\epsscale{1.0}
\plotone{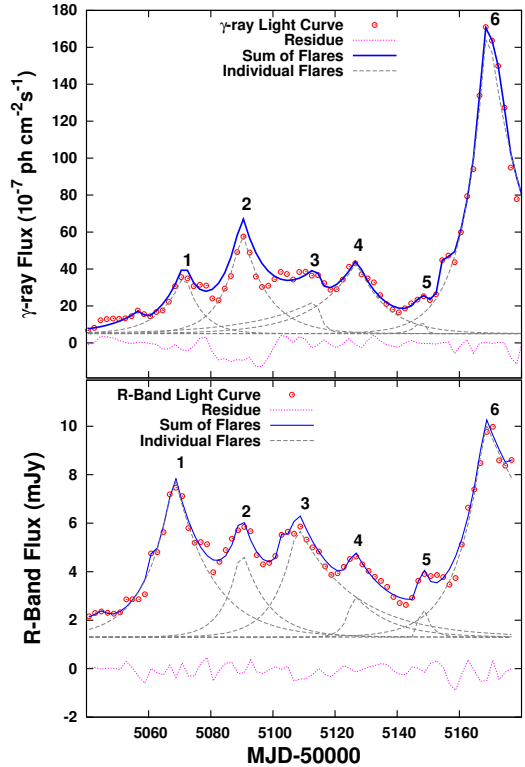}
\caption{Red open circles show smoothed $\gamma$-ray (0.1-300 GeV) and optical (R-band) light curves of the blazar 3C 454.3 during the second half of 2009. Blue solid curves correspond to summed flux after modeling the light curve as a superposition of six individual flares and a constant background level while magenta dotted lines denote the residual flux after subtracting the model flares, and the gray dashed lines indicate the underlying individual flares. Flares denoted by the same number in both panels peak within $\pm$3 days of each other and are probably physically related (``flare-pairs"). These flare pairs are listed in Table \ref{gammaop_area}.}
\label{gammaop_modelfit}
\end{figure}

To ensure that even small-timescale flares are resolved, we smooth the $\gamma$-ray and optical-IR light curves of the blazar 3C 454.3 during 2009 August--December with a Gaussian smoothing function of FWHM 2 days. Figure~\ref{gammaop_modelfit} compares the smoothed light curves with the summed flux (sum of contributions from all the model flares at all epochs). We identify 6 $\gamma$-ray/optical flare pairs in which the flux at both wavebands peaks at the same time within $\pm 3$ days. For each of the flares, we determine the time of the peak, width (defined as the mean of the rise and decay timescales), and area under the curve from the best-fit model. We calculate the area under the curve for each flare by integrating the photon flux over 0.1-300 GeV using average spectral index from the \textit{Fermi} 2-yr catalog \citep{abd11_2yrcatalog} and then integrating over the duration of the flare. This represents the total energy output of the outburst.  Table~\ref{gammaop_area} lists the parameters of each flare pair, along with the ratio, $\rho$, of $\gamma$-ray to optical energy output. 

\begin{figure}
\epsscale{1.0}
\plotone{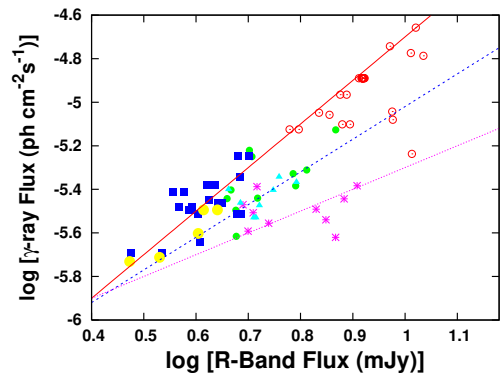}
\caption{The data points show the $\gamma$-ray vs. optical fluxes for each of the flares identified in Table \ref{gammaop_area} with each symbol denoting a different flare number in Figure \ref{gammaop_modelfit}: magenta asterisks-1, green filled circles-2, cyan triangles-3, blue filled squares-4, yellow filled circles-5, and red open circles-6. The lines denote three forms of the numerical relation between the fluxes in those two bands, e.g., $\rm F_{\gamma}\sim F_{\rm op}^x$, where x=1 (magenta dotted), 1.5 (blue dashed), and 2 (red solid).}
\label{gamma_vs_op}
\end{figure}

The presence of a corresponding R-band flare for each of the significant $\gamma$-ray flare implies that the emission at both wave bands is generated by the same electrons and hence provides strong support to the leptonic models \citep[e.g.,][]{bon09}. However, with some tuning, some hadronic models can also reproduce this behavior, so our results do not definitively rule out this alternative.
The total energy output of the $\gamma$-ray flares are larger than that in the optical by 1-2 orders of magnitude on average. This is consistent with the SED of this source \citep{bon11} which shows that it emits much greater energy at $\gamma$-ray-frequencies than that at the optical wave band. Table~\ref{gammaop_area} also shows that the energy output at the same energy band as well as the total energy output of the sum of a $\gamma$-ray flare and the corresponding optical flare varies significantly from one event to the other. Figure \ref{gamma_vs_op} shows the $\gamma$-ray flux vs. optical flux plot for each of the identified flare pairs. The lines denote three forms of the numerical relation between the fluxes in those two bands, e.g., $\rm F_{\gamma}\sim F_{\rm op}^x$, where x=1, 1.5, and 2.  

Exact location of the $\gamma$-ray production region in blazars is not well-determined. Optical-UV photons coming from the broad line region (BLR) may be inverse-Compton-scattered by the electrons in the jet, in which case the bulk of the observed luminosity appears in the $\gamma$-ray band \citep[e.g.,][]{ghi10}. But previous time variability studies from the EGRET era have indicated that the $\gamma$-rays are produced downstream of the very long baseline interferometry (VLBI) core (which lies $\gtrsim$1 pc from the SMBH) \citep{jor01,lah03}. $\gamma$-$\gamma$ absorption of very high energy $\gamma$-ray photons by the radiation field of the BLR may present another problem for a model of very high energy $\gamma$-ray emission inside the BLR of luminous quasars \citep[e.g.,][]{don03,rei07}. On the other hand, the external photon density at $\gtrsim$1 pc from the base of the jet is theoretically not large enough to generate the amount of $\gamma$-rays observed from many blazars \citep[e.g.,][]{ghi09,sik09}. 

\begin{deluxetable*}{ccccccccc}
\tablecolumns{7}
\tablewidth{0pt}
\tablecaption{Time, Total Energy Output (Area), and Temporal Width of $\gamma$-ray/Optical Flare Pairs Identified in the Light Curves of the Blazar 3C 454.3 During 2009 August-December.\label{gammaop_area}}
\tablehead{
\colhead{Flair Pair} &\multicolumn{3}{c}{$\gamma$-ray}&\colhead{} & \multicolumn{3}{c}{optical} & \colhead{$\rho$\tablenotemark{d}}\\
\cline{1-9}\\
\colhead{ID} &\colhead{Date\tablenotemark{a}} &\colhead{Area\tablenotemark{b}}&\colhead{Width\tablenotemark{c}}&\colhead{} & \colhead{Date\tablenotemark{a}} &\colhead{Area\tablenotemark{b}}&\colhead{Width\tablenotemark{c}}}
\startdata
1&   5071 &  180  &   6.5  &&   5068 &  5.6 & 10.5  &    32 \\ 
2&   5090 &  356  &   7.5  &&   5089 &  2.1 & 7.0   &  170 \\
3&   5112 &  164  &  11.0 &&   5109 &  4.7 & 12.8 &    35 \\
4&   5127 &  380  &  11.5 &&   5125 &  1.0 & 7.0   &  380 \\
5&   5148 &  13   &    1.8  &&   5147 &  0.2 & 1.5   &    65 \\
6&   5169 &  1084 &  10.5 &&   5168 &  6.5 &  18.3 & 167 \\
\enddata
\tablenotetext{a}{Date of peak in units of MJD-50000.}
\tablenotetext{b}{Units: $10^{-5}$ erg cm$^{-2}$.}
\tablenotetext{c}{Mean of rise and decay timescale in units of days.}
\tablenotetext{d}{Ratio of $\gamma$-ray to optical energy output integrated over flare.}
\end{deluxetable*}

Recently, \citet{agu11} have shown that bright $\gamma$-ray flares in the jet of the blazar OJ 287 occur $>$14pc from the central engine. They find this by analyzing a combination of time-dependent multi-waveband flux and linear polarization observations, and sub-milliarcsecond-scale polarimetric VLBI images at 7 mm. Similar location ($\gtrsim$few pc from the black hole) of the production of $\gamma$-ray emission has been inferred by \citet{mar08} and \citet{mar10} as well for the blazars BL Lac and PKS 1510-089, respectively. They argue that large multi-waveband outbursts are triggered by the interaction of moving plasma blobs with a standing shock present in the jet seen as the ``core" in the pc-scale jet in Very Long Baseline Array (VLBA) images. In all these cases, the large multi-frequency flares were preceded by smaller flares caused by the movement of those plasma blobs along the jet axis. The blobs follow a helical path at the acceleration-collimation zone of the jet due to the nature of the magnetic field present there. That causes the movement of the blobs to align with our line of sight and hence their emission to be Doppler boosted during some intervals.

To understand the location and mechanism of the $\gamma$-ray and optical emission in this blazar we discuss the dependence of the observed synchrotron (F$_{\rm synch}$), synchrotron self-Compton (F$_{\rm SSC}$), and external-Compton (F$_{\rm EC}$) flux on three relevant parameters, namely, total number of emitting electrons (N${\rm _e}$), magnetic field (B) and Doppler factor ($\delta$). We choose these parameters because these are independent of each other in the following functions:\\
\begin{equation}
\mathrm{F_{synch}\sim N_e B^{1+\alpha_O} \delta^{3+\alpha_O}}
\end{equation}
\begin{equation}
\mathrm{F_{EC}\sim N_e \delta^{4+2 \alpha_g} U^{\prime}_{ext}}
\end{equation}
\begin{equation}
\mathrm{F_{SSC}\sim N_e \delta^{3+\alpha_g} U^{\prime}_{synch}},
\end{equation}
where $\alpha_O$ and $\alpha_g$ are the spectral indices of the synchrotron and IC emission respectively, U$^{\prime}_{\rm ext}$ is the external seed photon field, and U$^{\prime}_{\rm synch}$ is the same due to synchrotron emission in the jet itself. The observed synchrotron emission from moving plasma in a relativistic jet is amplified by a factor $\delta^{3+\alpha}$ \citep{urr95}. In the case of EC emission the amplification factor is $\delta^{4+2\alpha}$ due to the additional Lorentz transformation between the seed photon and jet rest frame \citep{der95}. There is no additional factor of $\delta$ in F$_{\rm SSC}$ in Equation 5 because the synchrotron seed photons are also from the jet rest frame where the emitting particles reside. We note that i) N$_e$ is the number of electrons contributing to the observed flux and is related to the normalization (N$_0$) of the electron spectrum. The actual dependence of flux is on N$_0$. If the minimum energy of the electron distribution ($\gamma_{\rm min}$) remains unchanged N$_e$ is proportional to N$_0$, and ii) Above mentioned fluxes are at fixed observed energy bands as shown in Figure \ref{gamma_vs_op}. 
There are other relevant quantities such as the size of the emitting region but we assume that it remains constant in order to determine the effect of the above parameters on the observed flux. 
A detailed numerical calculation is required to accurately model the effect of various quantities on the observed fluxes but such a calculation is beyond the scope of the paper \citep[see][for details]{tra09,der95}. Our goal here is to compare the observed data to approximate theoretical scenarios. 

From Equation 3 and 5, we get:\\
\begin{equation}
\mathrm{F_{SSC}\sim N_e^2 B^{1+\alpha_O} \delta^{3+\alpha_g}}
\end{equation}

Therefore, i) If the variation is due to a change in B, F$_{\rm EC}$ will not vary while F$_{\rm synch}$ $\sim$$B^{1+\alpha_O}$, i.e., the $\gamma$-ray and the optical variation will not be correlated. On the other hand, F$_{\rm SSC}$$\sim$F$_{\rm synch}$ ii) If it is due to a variation in N$_e$, F$_{\rm EC}$$\sim$F$_{\rm synch}$ and F$_{\rm SSC}$$\sim$F$_{\rm synch}^2$ and iii) If the variation is due to a change in $\delta$, F$_{\rm EC}$$\sim$ F$_{\rm synch}^{(4+2\alpha_g)/(3+\alpha_O)}$ and F$_{\rm SSC}$$\sim$ F$_{\rm synch}^{(3+\alpha_g)/(3+\alpha_O)}$. During the interval of these flares, i.e., MJD 55060 to 55160, OIR spectral index from SMARTS light curves is 1.55$\pm$0.05. The $\gamma$-ray spectral index during the same interval is 1.5$\pm$0.1 \citep{ack10_flares}. For $\alpha_{\rm O}=1.55$ and $\alpha_{\rm g}=1.5$, F$_{\rm EC}$ $\sim$ F$_{\rm synch}^{1.5}$ and F$_{\rm SSC}$ $\sim$ F$_{\rm synch}^{1.0}$. Therefore, from Figure \ref{gamma_vs_op} we can confirm that if the $\gamma$-rays are EC in nature, the variation is not due to a change in B. No other possibilities can be ruled out. Other relations between the $\gamma$-ray and optical flux variation is possible if the variation is due to a combination of two or more of the above scenarios.

It is possible that the large multi-frequency flare of 3C 454.3 during 2009 December is another case where a large outburst is generated due to the interaction of a moving plasma knot with a standing shock present in the jet. This can be tested by following the variation in the structure of the pc-scale jet with VLBA imaging for a sufficient duration. Such a study\footnote{http://www.bu.edu/blazars/VLBAproject.html} indeed shows that a new knot (moving plasma blob) was coincident with the core (standing shock) on 2009 November 27 ($\pm$15 Days)(Alan Marscher, private communication). This implies that the large $\gamma$-ray and optical outbursts might take place in the jet near the VLBA core located at $\sim$18 pc from the central engine \citep{jor10}.  

In that case, the smaller flares preceding the large outburst may be due to Doppler boosting of the emission from the same knot when it is propagating through the acceleration-collimation zone of the jet and getting in and out of our line of sight. This is consistent with the following properties of these flares: i) We can see in Figure \ref{gamma_vs_op} that the $\gamma$-ray vs. optical flux data points for four out of five smaller flares (yellow, green, cyan, and magenta) are roughly consistent with $\rm F_{\gamma}\sim F_{\rm op}^{1.5}$ or $\rm F_{\gamma}\sim F_{\rm op}^{1.0}$, as expected if the variation is due to changes in the Doppler factor. ii) The range of F$_{\gamma}$ values on Figure \ref{gamma_vs_op} corresponding to the five flares preceding the large flare around MJD 55169 denoted by flare number 6 in Figure~\ref{gammaop_modelfit} correspond to a change by a factor of $\sim$3.7 while that for F$_{\rm op}$ is $\sim$2.8. The fractional change required in $\delta$ to account for these changes is 20-25\% which is definitely plausible \citep{jor05} and finally iii) These flares were near-simultaneous at all wave bands and the recurrence time of the peaks were similar. The first peak is on MJD 55071 and after that each flare is $\sim$20 days after the previous flare within $\pm$3 days. This can be explained by the movement of the plasma blob through a helical path while its emission is affected by Doppler boosting at regular intervals \citep{cam92,mar08,mar10,agu11}.

The scenario described above is not the only possible process to explain the above observations. As described above, the variation could be due to changes in N$_e$, $\delta$ or a combination of one or both of them with changes in B. The physical reason for a change in N$_e$ or B is not clear from the data. For example, the acceleration of electrons due to the presence of instability in the jet could increase N$_e$. Alternatively, stochastic injection of energy in the jet plasma from the central engine and its dissipation at different distances from the base of the jet \citep{bonnoli11} could explain the observations as well.
 
If the Doppler beaming explanation is true, this adds to the increasing evidence that at least in some blazars, copious $\gamma$-ray emission is produced farther down the jet and in those cases the BLR cannot be a significant source of seed photons which are inverse-Compton scattered to $\gamma$-ray energies by the energetic electrons in the jet. \citet{tav10_location} showed that during this high state in 2009 December, the $\gamma$-ray flux from 3C 454.3 changed by a factor of $\sim$5 on a 6--12 hr timescale. This implies that the size-scale of emission region, obtained from the variability timescale ($\sim$0.01 pc) is smaller than the approximate cross-section of the jet of 3C 454.3 at 18 pc ($\sim$0.05 pc, assuming $\theta_{\rm jet}$$\sim$0.2$^{\circ}$ from \citealt{jor05}). This may imply that the size-scale of the $\gamma$-ray emission regions is smaller than the diameter of the local cross-section of the jet. Both of these provide strong constraints to the theoretical models of $\gamma$-ray production in the jets of blazars.

Stronger constraints on the location and mechanism of emission will come from a detailed SED modeling of the SMARTS, Fermi and other multi-wave band data of the above individual successive flares, to be addressed in a future paper.

\section{Summary and Conclusions}
This paper presents the time variability properties of a sample of six $\gamma$-ray-bright blazars at optical-IR frequencies as well as $\gamma$-ray energies: AO 0235+164, 3C 273, 3C 279, PKS 1510-089, PKS 2155-304, and 3C 454.3. The light curves were obtained as part of the Yale/SMARTS program in 2008-2010 to monitor all bright southern Fermi-LAT-monitored blazars on a regular cadence, at optical and near-infrared (BVRJK) wave bands. 
Our main conclusions are as follows:\\
(1) We find the optical-IR variability properties to be remarkably similar to those at $\gamma$-ray energies. This is consistent with the general picture of the leptonic model where the lower (optical-IR) and higher ($\gamma$-ray) energy emission is generated by the same population of electrons through synchrotron and inverse-Compton processes, respectively. However, more rapid variations indicate that the electrons producing $\gamma$-ray emission in the blazar 3C 279 may have slightly larger energy than those generating optical-IR radiation in this source. We note that some hadronic models are also able to reproduce the similarity of the optical-IR variability properties of blazars with those at the $\gamma$-ray bands and hence we cannot rule out those models on this basis.  \\
(2) The discrete auto-correlation function (ACF) of the variability of these six blazars at optical-IR and $\gamma$-ray wave bands do not show any periodicity or characteristic timescale. The shape and width of the ACFs are very similar in all bands in the 3 sources where the temporal sampling at both energies are identical. This indicates that the emission regions are of similar sizes with light-crossing time being the dominant timescale. \\
(3) The power spectral density functions of the R-band light curves of all six blazars are fit well by simple power-law functions with negative slopes, implying there is higher amplitude variability on longer than on shorter timescales. No evidence of a characteristic timescale, including a break, is identified in the PSD of any of the sources. The slope of the average PSD of the sample of six blazars is similar to what \citet{abd10_timing} found for the $\gamma$-ray variability of a larger sample of bright blazars. This is consistent with the conclusion in item (1) that the lower (optical-IR) and higher ($\gamma$-ray) energy emission is generated by the same population of electrons.\\
(4) For all six sources, the shortest timescales over which the $\gamma$-ray flux changes by a factor of 2 is $\lesssim$1 day. For the blazars AO 0235+164 and PKS 1510-089, this timescale in optical-IR energies is $1-2$ days, consistent with those in the $\gamma$-ray band. However, in 3C 454.3, the doubling timescale is longer at shorter wavelength optical-IR bands. This may be due to the disk contribution in the bluer bands which dilutes the jet variability and hence increases the doubling timescale. In 3C 273 the flux does not change by a factor of 2 during this observing program which may also be a result of significant contribution from the disk in the optical-near IR emission of this object. 
\\
(5) The variability properties, including the ACF peak value, the relative nature of the short and long-term variability at $\gamma$-ray and optical wave bands, and the doubling timescales are different in the BL Lac 2155-304 from the FSRQs in our sample. This may be caused by: i) the $\gamma$-rays in PKS 2155-304 are dominated by the SSC process unlike the FSRQs in which it is dominated by the EC contribution and ii) the electrons generating the $\gamma$-ray emission in this blazar are at a higher energy than those producing the optical-IR emission unlike the FSRQs where the GeV and R-band emission is produced by electrons of similar energies.  \\
(6) We decompose the optical-IR and $\gamma$-ray light curves to investigate the properties of the individual flares present in those light curves. The prominent flares present in the optical-IR as well as the $\gamma$-ray light curves of these blazars are predominantly symmetric, i.e., have similar rise and decay timescales. This indicates that the the long-term variability is dominated by the crossing time of radiation or a disturbance through the emission region and not by the energy-loss time of the emitting electrons due to radiation losses. \\
(7) Finally, we perform a detailed decomposition of the SMARTS and Fermi light curves of 3C 454.3 during and before its large outburst in 2009 December to understand how and where the emission in the prominent outbursts of blazars is generated. The total energy output, the ratio of $\gamma$-ray to optical energy output, and the $\gamma$-ray vs. optical flux relation of six individual flares of the blazar 3C 454.3 during 2009 August to December vary significantly from one event to the other.\\
(8) In 3C 454.3, the smaller flares at both optical-IR and $\gamma$-ray wave bands during 2009 August-November, and the subsequent large multi-frequency outburst in 2009 December, might be triggered by the movement of a plasma blob along a helical path in the jet and its subsequent interaction with a standing shock present in the jet, respectively. If this scenario is right, the location of the large $\gamma$-ray outburst in 2009 December is identified at $\sim$18 pc down the jet from the central engine, where a strong radio-emitting component is observed. This would pose strong constraints on the models of high energy emission in the jets of blazars.


\section{Acknowledgments}
RC received support from Fermi GI grant NNX09AR92G. SMARTS observations of LAT-monitored blazars are supported by Fermi GI grant 011283 and NNX10A043G. CDB, MMB and the SMARTS 1.3m observing queue also receive support from NSF grant AST-0707627. GF is supported by Fermi GI grant NNX10A042G. JI is supported by the NASA Harriet Jenkins Fellowship program and NSF Graduate Research Fellowship under Grant No DGE-0644492.


\end{document}